Nano dimensional hybrid organo-clay Langmuir-Blodgett films

Syed Arshad Hussain[*], S. Chakraborty, D. Bhattacharjee

Department of Physics, Tripura University, Suryamaniangar – 799022, Tripura, India

* Corresponding author
Email: sa_h153@hotmail.com, sah.phy@tripurauniv.in
Ph: +919862804849 (M), +91381 2375317 (O)
Fax: +913812374802

**Abstract:**

Clay mineral particles are interesting nanosized building blocks due to their high aspect ratio and the chemical properties. The main interest in this nanosized building blocks results essentially from the colloidal size and the permanent structural charge of the particles. Smectites or swelling clay minerals are naturally occurring nanomaterials that can be fully delaminated to elementary clay mineral platelets in dilute aqueous dispersion. This dilute aqueous smectite suspensions are well suited to convert into functional nanofilms. The functionalization is performed by ion exchange reaction with amphiphilic molecules carrying the desired functionality, such as chirality, two photon absorption, energy transfer, optical nonlinearity and magnetism, which are due to the nature of the amphiphilic cations and to the organization of both the amphiphilic molecules and the elementary clay mineral platelets. Controlling the structure of materials at the nanometre scale is of fundamental importance for tailoring such materials properties. Langmuir Blodgett films are known for a high degree of organisation of organic molecules. Fundamental research into the organization of molecules at clay mineral platelets is necessary to optimize the materials for specific applications. This paper gives an overview of organo-clay hybrid Langmuir-Blodgett nano films.

# 1. INTRODUCTION

Worldwide interest in nanomaterials is rapidly increasing due to their unique properties. Nanostructured materials have gained widespread interest because of their potential applications in a number of diverse technological fields. In hybrid nanostructured organic-inorganic ultrathin films the organic materials impart flexibility and versatility whereas, the inorganic clay particles provide mechanical strength and mechanical stability and can have unique conducting, semiconducting or dielectric properties. Construction and study of such organic-inorganic nanostructured materials is an important target for the researchers for developing functional materials such as sensors, electrode-modifiers, nonlinear optical devices and pyroelectric materials.

Recently, the development of nanoscience and nanotechnology ignited a new area of interest of clay minerals. Based on their nano-sized layers as well as the nano-sized interlayer space, clay minerals can act as naturally occurring nanomaterials or as nano-reactors for fabrication of nano-species, nanoparticles or nanodevices [1]. Clay mineral particles are interesting nanosized building blocks due to their high aspect ratio and the chemical properties. The main interest in this nanosized building blocks results essentially from the colloidal size and the permanent structural charge of the particles. One of the outstanding properties of the clay is the simultaneous incorporation of polar or ionic molecules into the interlamellar spaces (intercalation) resulting in hybrid materials. The property of intercalation makes it easy to prepare the composite materials. Clay particles provide a host medium to assemble and organise guest molecules. By studying and adjusting the host, guest and intercalation reaction materials can be designed with unique structures, novel chemical and physical phenomena, and enhanced mechanical properties. If the orientation of the incorporated molecules can be controlled, the clay composite materials would be applicable to devices for current rectifying, nonlinear optics, and one-way energy transfer etc.

Controlling the structure of materials at the nanometer scale is of fundamental importance for tailoring the materials properties. Langmuir-Blodgett (LB) films are known for a high degree of organization of organic molecules [2-10]. Recently, LB technique has been extended to the preparation of hybrid organo-clay mono- and multilayers [11-15]. With the LB technique hybrid films can be prepared, consisting of a two – dimensional ordering of single clay mineral layers, hybridized with amphiphilic cations. With a suitable choice of cations films can be prepared, which show nonlinear optical (NLO) properties, two photon absorption (TPA), J dimer and fluorescence resonance energy transfer (FRET) etc. In this article we provide an overview of the organo-clay hybrid films prepared by LB technique. Also few interesting properties of organo-clay hybrid films prepared in our laboratory have been demonstrated.

# 2. CLAY MINERALS

Clay has been known to, and used by, humans since antiquity. Human beings found various applications of layered clay minerals since prehistoric civilization due to their widespread distribution and a great diversity of reactions in nature. Clay minerals are a class of phyllosilicates which usually form as a result of chemical weathering of other silicate minerals at the surface of the earth. These layered minerals consisting of stacks of negatively charged two-dimensional aluminosilicate layers. Each layer is comprised of fused sheets of octahedra of $Al^{3+}$, $Mg^{2+}$ or $Fe^{3+}$ oxides and sheets of tetrahedra of $Si^{4+}$ oxides. These tetrahedral and octahedral sheets can be arranged in clay layers in a variety of different ways.The most important clay minerals, both industrially and scientifically, are the 1:1 (caly minerals made up of one octahedral and one tetrahedral sheet, and sometime denoted as T-O clay) clay minerals and the 2:1 (clay mineral combines two tetrahedral sheets which sandwich one central octahedral sheet,denoted as T-O-T clay) clay minerals with moderate charge, also called swelling clays or smectites. Smectite clays are nanomaterials.

Fig 1 shows the structure of basic 2:1 clay minerals. The elementary platelets are a few tenths to a few hundreds of nanometers wide and long and 0.96 – 1.50 nm thick.Tthe exact thickness depends on the number of adsorbed water layers. The elementary platelets or individual clay layer carry a negative surface charge,which is the sum of variable and permanent charges. The variable charge occurs at the edges of clay layers mainly at hydroxyl groups, which exhibit acid/base properties; whereas the permanent charge is due to the presence of non-equivalent, isomorphic substitution of the central metal atoms within the octahedral and/or tetrahedral sheets. For semectites, the value of the permanent charge is always significantly higher than that of the variable charge. The typical negative charge per $O_{10}(OH)_2$ unit is 0.25 – 0.60 e (e = $1.66 \times 10^{19}$ C). This charge is compensated by cations which exist in the region between the clay layers known as the interlayer region. These interlayer cations are exchangeable cations and may exchange their places with other cations under appropriate conditions. The corresponding charge-neutralizing and exchangeable cations together with one or two water layers are located in the interlamellar space between the elementary platelets. A group of elementary platelets forms a clay particle. If such a particle is immersed in water, water molecules are attracted into the interlamellar spaces and the clay particle swells. Ultimately, the aqueous clay suspension consists of randomly-oriented, more or less freely moving, elementary clay platelets. In any case, the degree of swelling depends on the charge density of the clay minerals, size and shape of the elementary platelets and particles, the type and charge of the exchangeable cation and the chemical and thermal history of the clay mineral sample. Thus, one can transform a clay particle, consisting of a number of more or less well-oriented elementary platelets, into an aqueous suspension with randomly moving, elementary clay platelets. In this state the charge compensating cations are exchangeable with almost any type of cation, be it inorganic, organic or organometallic. This property is the basis for the idea that the elementary platelets of smectites can be converted into nanoparticles with preset properties. One needs (1) an exchangeable cation with the desired property (light absorption, light emission, redox, acid, base); (2) a specific organization of the exchangeable cation at the surface of the elementary clay platelets; (3) the organization of the elementary platelets into monolayers and multilayers. This means that a double organization is necessary: (i) organization of the molecules at the surface, and (ii) organization of the elementary clay platelets, carrying the desired molecules.

The development of smectites into nanomaterials with specific functionalities is far less advanced. This development requires knowledge of (i) the organisation of the smectite nanoparticles in two and three dimensions; (ii) the organisation of molecules at the surface of these particles; (iii) the introduction of functionalities in the nanofilms either via the clay mineral particles or via the adsorbed molecules. Up to now functionalities (nonlinear optics, magnetism, chirality) have been introduced with a proper choice of the adsorbed molecules and their organisation. In any case, with such nano films one can study the fundamental properties of the elementary particles themselves at the level of the individual particle in the nanofilms and those of the adsorbed molecules with an unprecedented precision and depth.

## 3. NANO CLAY PARTICLES IN DRUG DELIVERY SYSTEM

Clay minerals are naturally occurring inorganic cationic exchangers which may undergo ion exchange with basic drugs in solution. Swelling clays or smectites, especially montmorillonite and saponite, have been the more commonly studied because of their higher cation exchange capacity compared to other pharmaceutical silicates (such as talc, kaolin and fibrous clay minerals) [16]. There are also several mechanisms which involved in the interaction between clay minerals and organic molecules [17]. Layered materials which can accommodate polar organic or biomolecular compounds have received more attention in this recent years as a drug delivery vechicle. Recently there are many reviewed article on nanoclay drug delivery system [16,18]. Nano-clay drug delivery systems are widely used for those drugs which require the slow release rate. Calcium montmorillonite has also been used extensively in the treatment of pain, open wounds, colitis, diarrhea, hemorrhoids, stomach ulcers, intestinal problems, acne, anemia, and a variety of other health issues. Clay minerals may also be used typically as antiseptics, disinfectants, dermatological protectors, anti-inflammatories, local anesthetics and keratolytic reducers [19].

## 4. ORGANO – CLAY HYBRİD FİLM PREPARATİON

Organo-clay hybrid films can be prepared by dropcasting, spincoating, layer-by-layer (LbL) self assembly and the Langmuir–Blodgett (LB) technique. LbL and LB techniques provide a better platform to prepare organo-clay hybrid films. However, the LB technique is preferred because it leads to smoother films with a good organization of elementary clay sheets and organic molecules. Also the LB technique is very flexible where by changing and controlling various parameters it is possible to control the organization as a whole [9,10,20-22]. LB film is prepared at the gas-liquid interface where the hybrid monalayer can be compressed and this leads to additional control on the lateral distribution of the clay mineral particles and molecules.

In the following sections some basic preliminaries about LB technique and organo-clay hybrid film formation using LB technique have been discussed.

## 5. LANGMUIR–BLODGETT TECHNIQUE
### 5.1. LB-compatible materials

In order to form a Langmuir monolayer, it is necessary for a substance to be water insoluble and soluble in a volatile solvent like chloroform or benzene. LB compatible materials consist of two fundamental parts, a "head" and a "tail" part. The "head" part is a hydrophilic (water loving) chemical group, typically with a strong dipole moment and capable of hydrogen bonding, like OH, COOH, $NH_2$, etc. The "tail" part on the other hand is hydrophobic (water repealing), typically consisting of a long aliphatic chain. Such molecules, containing spatially separated hydrophilic and hydrophobic regions, are called amphiphiles [23]. Typical examples of LB compatible materials are the long chain fatty acids (stearic acid, arachhidic acid etc.) and their salts. A schematic of LB compatible molecule is shown in fig. 2a. If amphiphile molecules arrive at the air-water interface with their hydrophobic tails pointing towards the air and hydrophilic group towards water, the initial high energy interface is replaced by lower energy hydrophilic-hydrophilic and hydrophobic-hydrophobic interfaces, thus lowering the total energy of the system. Hence, the molecules at the interface are anchored, strongly oriented normal to the surface and with no tendency to form a layer more than one molecule thick.

**5.2. Langmuir monolayer formation**

Fig-2b shows a typical LB film deposition instrument (designed by Apex Instrument Co, India) installed in our laboratory and fig-3 shows the schematic of the LB technique. Essentially all LB film works begin with the Langmuir-Blodgett trough, or Langmuir Film balance, containing an aqueous subphase (Fig. 2b). Moveable barriers that can skim the surface of the subphase permit the control of the surface area available to the floating monolayer. Nowadays sophisticated Langmuir-Blodgett (LB) film deposition instruments are designed and marketed by several companies. To form a Langmuir monolayer film, the molecules of interest is dissolved in volatile organic solvents (chloroform, hexane, toluene, etc.) that will not dissolve or react with the subphase. The dilute solution is then minutely placed on the subphase of the LB trough with a microliter syringe [fig. 3a]. The solvents evaporate quickly and the surfactant molecules spread all over the sub phase in the LB trough. In order to control and monitor the surface pressure, $\pi$ (this quantity is the reduction of surface tension below that of clean water), the barrier intercepts the air-water interface is allowed to move so as to compress or expand the surface of the film. Wilhelmy plate arrangement is used to measure the surface pressure. In this method a small piece of hydrophilic material, usually a piece of filter paper, intercepting the air-water interface and is supported from the arm of an electronic microbalance which is interfaced with a computer is used. The force exerted is directly proportional to the surface tension.

There are several techniques available to monitor the state of the floating monolayer. The measurement of surface pressure ($\pi - A$) as a function of area per molecule (A) in the monolayer film is known as isotherm characteristics. This characteristic is easily obtained and contains much useful information about the mono-molecular films at the air-water interface [3, 5, 8, 24, 25]. A conceptual illustration of the surface pressure versus area per molecule isotherm is shown in Fig. 4. As the pressure increases, the two-dimensional monolayer goes through different phases that have some analogy with the three-dimensional gas, liquid, and solid states. If the area per molecule is sufficiently high, then the floating film will be in a two-dimensional gas phase where the surfactant molecules are not interacting. As the monolayer is compressed, the pressure rises, signaling a change in phase to a two-dimensional liquid expanded (LE) state, which is analogous to a three-dimensional liquid. Upon further compression, the pressure begins to rise more steeply as the liquid expanded phase gives way to a condensed phase, or a series of condensed phases. This transition, analogous to a liquid-solid transition in three dimensions, does not always result in a true two-dimensional solid. Rather, condensed phases tend to have short-range structural coherence and are called liquid condensed (LC) phases. If the surface pressure increases much further the monolayer will ultimately collapse or buckle, still not being a single molecule in thickness everywhere. This is represented by a sudden dip in the surface pressure as the containment area is decreased further, such as is shown in Fig. 4.

**5.3. Langmuir-Blodgett films**

The term "Langmuir-Blodgett film" traditionally refers to monolayers that have been transferred from the water sub-phase onto a solid support. The substrate can be made of almost anything. However the most common choices are glass, silicon, mica, quartz, etc. Vertical deposition is the most common method of LB transfer [20, 26, 27]. In principle the Langmuir-Blodgett deposition method simply consists of dipping and pulling a solid substrate, orientated vertically, through the floating mono-layer while keeping the surface pressure constant at a desired value (Fig. 3). It is also a common practice to coat the substrate with a highly hydrophobic or hydrophilic material. The rate at which the substrate is dipped or pulled through the monolayer must also be precisely controlled and kept constant at a very low value (typically 5 mm/min) [3, 5, 7, 24]. The surface pressure for film deposition is normally chosen to be in the solid-like region. However, at any pressure film can be deposited. The transfer of monolayer film occurs via hydrophobic interactions between the alkyl chains and the substrate surface or the hydrophilic interaction between the head groups of the molecules and the hydrophilic substrate surface. Subsequent dipping or pulling deposits a second layer on top of the first, the process simply being repeated until the desired number of layers has been deposited (fig. 3). LB films have been explored for applications that include electronics, optics, microlithography, and chemical sensors, as well as biosensors or biochemical probes [10, 20, 26, 27].

**5.4. Hybrid film formation using LB technique**

A highly organized layer-by-layer deposition of elementary clay mineral particles can be achieved with the LB technique. In early days hydrophobic clay minerals are dispersed in a volatile organophilic solvent such as chloroform and this dilute suspension is spread over the water surface in a LB trough and after allowing sufficient time for the chloroform to evaporate the film of the hydrophobic clay mineral is compressed and transferred to solid substrate [28, 29].

A more elegant method is to spread the amphiphilic cations on the water surface of a dilute aqueous dispersion [11-13, 30, 31]. It is possible to prepare aqueous dispersion consisting of nanometer thick single clay platelets. The clay is dispersed in water and stirred in a magnetic stirrer for 24 hours followed by ultrasonication for 30 minutes before use. In the process of hybrid film formation the Langmuir–Blodgett (LB) trough is filled up with the dilute aqueous dispersion of a smectite with Na+ or Li+ as exchangeable cations. The small univalent cations ensure maximal swelling and the presence of elementary clay mineral platelets of 0.96 nm thickness (not counting the hydration water). A solution of an amphiphilic cation in a volatile organic solvent is spread at the air–water interface of the LB trough [fig. 5]. The solvent evaporates and a monolayer of molecules is formed. A rapid ion exchange reaction takes place between the smectite particles and the amphiphilic cations, leading to a hybrid monolayer at the air–water interface, consisting of one layer of amphiphilic cations and, ideally, one layer of elementary smectite sheets of 0.96 nm thickness. This hybrid monolayer can be compressed and surface pressure–area isotherms are constructed. At a fixed surface pressure the hybrid clay–organic monolayer is transferred to a substrate either by vertical deposition in upstroke or by horizontal deposition. Surfactants for Langmuir–Blodgett clay films include alkylammonium cations, organic dyes and metal ion complexes [32, 33]. Once the LB film is prepared it can be characterized by different techniques in order to check their possible functionality.

## 6. ORGANIZATION OF CLAY MINERAL SHEETS

The core of nanofilm research required the thorough knowledge of organization of molecules. In nanofilms, molecules may be manipulated in such a way that the organized molecular system may have different specific properties which are absent in randomly oriented molecular system. It has been known for long time that molecules can adopt specific organizational structure in the interlamellar spaces of the smectite depending on their shape, size and charge of the molecules through one of the outstanding properties of the clay known as intercalation. The property of intercalation makes it easy to prepare the composite materials. If the orientation of the incorporated molecules can be controlled, the clay composite materials would be applicable to devices for current rectifying, nonlinear optics, and one-way energy transfer etc.

Two levels of organizations exist in these hybrid monolayers: the organization of the clay particles in a monolayer and the organization of organic molecules adsorbed on the clay surface. Spectroscopy has revealed the organization of the dye molecules at the clay-mineral surfaces. Monomers and different types of aggregates are found, depending on (1) the charge density of the clay mineral; (2) the particle-size distribution of the clay minerals; (3) the nature of the exchangeable cations; (4) the degree of swelling in water; and (5) the nature of the dye, i.e. their specific chemical structure and their orientation at the clay-mineral surfaces in the presence and absence of water molecules.

Photochemical intercalation compounds with such functions as photocatalysis, energy storage, photoluminescence, photochromism, photochemical hole burning and nonlinear optics have already been reported [34-36]. In the hybrid clay mineral-dye systems the photofunction is provided by the dye molecules. The clay mineral acts as a host, which provides a two-dimensional organization of the dye molecules. This organization is essential to obtain a measurable photofunction.

For nonlinear optics, a non-centrosymmetric arrangement of the molecules at the clay-mineral surface is essential. Such is the case for LB films containing hybrids of clay mineral layers and non-amphiphilic bipyridyl and phenanthroline complexes [37- 39] and clay mineral layers, hybridized with zwitterionic compounds and cyanine dyes [40-42]. With MTTPB-clay mineral systems two-photon absorption (TPA) can be realized [43-45]. To obtain a reproducible signal, a low scattering material is necessary. This can be realized using synthetic clay minerals with particle sizes in the range 20–40 nm, significantly below the wavelength of light used in the experiments. Scattering is also reduced by using a mixture of water and dimethylsulfoxide as solvent.

## 7. INFLUENCE OF NANOCLAY LAPONITE ON FLUORESCENCE RESONANCE ENERGY TRANSFER

FRET is the relaxation of an excited donor molecule D by transfer of its excited energy to an acceptor molecule A. Consequently, the later emits a photon. The non-radiative energy transfer occurs as a result of dipole–dipole coupling between the donor and the acceptor, and does not involve the emission and reabsorption of photons. The process can be represented as

$$D + h\upsilon \rightarrow D^*$$
$$D^* + A \rightarrow D + A^* \quad [D \rightarrow \text{donor}, A \rightarrow \text{Acceptor}]$$
$$A^* \rightarrow A + h\nu$$

where h is the Planck's constant and $\nu$ is the frequency of the radiation. The two main conditions for FRET to occur are (i) sufficient overlap between the absorption band of acceptor (A) and the fluorescence band of donor (D) and (ii) the molecule A must be in the neighborhood of the molecules D. The efficiency of FRET depends on the inverse sixth power of the distance of separation between the donor and acceptor [46, 47]. Energy transfer manifests itself through decrease or quenching of the donor fluorescence and a reduction of excited state lifetime accompanied also by an increase in acceptor fluorescence intensity. Considering the proximity of the donor and acceptor molecules, FRET gives an opportunity to investigate and estimate distances between the two molecules. The relative orientations of the donor and acceptor transition dipoles also effect the FRET process.

Resonance energy transfer is not sensitive to the surrounding solvent shell of a fluorophore, and thus, produces molecular information unique to that revealed by solvent-dependent events, such as fluorescence quenching, excited-state reactions, solvent relaxation, or anisotropic measurements. The major solvent impact on fluorophores involved in resonance energy transfer is the effect on spectral properties of the donor and acceptor. Non-radiative energy transfer occurs over much longer distances than short-range solvent effects and the dielectric nature of constituents (solvent and host macromolecule) positioned between the involved fluorophores has very little influence on the efficiency of resonance energy transfer, which depends primarily on the distance between the donor and acceptor fluorophore.

The technique of FRET, when applied to optical microscopy, permits to determine the approach between two molecules within several nanometers. FRET is being used more and more in biomedical research and drug discovery today [48]. FRET is one of few tools available for measuring nanometer scale distances and the changes in distances, both in vitro and in vivo. Due to its sensitivity to distance, FRET has been used to investigate molecular level interactions. Recent advances in the technique have led to qualitative and quantitative improvements, including increased spatial resolution, distance range and sensitivity.

Energy transfer between two laser dyes N, N-dioctadecyl thiacyanine perchlorate (NK) and a octadecyl rhodamine B chloride (RhB) has been observed [11]. It was observed that presence of nanoclay sheet laponite increases the energy transfer efficiency. These two dyes NK and RhB are in principle suitable for FRET. Both the dyes are highly fluorescent. The fluorescence spectrum of NK sufficiently overlaps with the absorption spectrum of RhB.

Figure 7 shows the fluorescence spectra of pure NK, RhB in LB monolayer film and their mixture (NK:RhB = 50:50 volume ratios) in presence and absence of nanoclay sheet laponite. The excitation (absorption) wavelength ($\lambda_{ex} = 430$ nm) was selected approximately to excite the NK molecules directly and to avoid or minimize the direct excitation of the RhB molecules.

Figure 7 revels strong prominent NK fluorescence band where as the RhB emission band is very less in intensity in case of individual pure dye in LB films. The less intensity of pure RhB fluorescence band indicates very small contribution of direct excitation of the RhB molecules. The fluorescence spectra of NK-RhB mixture is very interesting, here the NK fluorescence intensity decreases in favour of RhB emission band. In this case the NK emission decreases due to the transfer of energy from NK to RhB. This transferred energy excites more RhB molecules followed by light emission from RhB, which is added to the original RhB fluorescence. As a result the RhB fluorescence intensity gets sensitized. Inset of figure 7 shows the excitation spectra measured with emission wavelength fixed at RhB (600 nm) fluorescence maximum in case of NK-RhB mixed films. Interestingly the excitation spectra possess characteristic absorption bands of NK. This confirms that the RhB fluorescence is mainly due to the light absorption by NK and corresponding transfer to RhB monomer. Thus FRET occurs from NK to RhB.

The most interesting observation in presence of nanoclay was that the NK fluorescence intensity decreases farther in favour of RhB fluorescence intensity (figure 7), resulting an increase in FRET efficiency.

Analysis of fluorescence spectra (figure 7) reveal that the spectral overlapping integral J(λ) between the fluorescence spectrum of NK (donor) and RhB (acceptor) absorption spectrum increases from $1.29 \times 10^{13}$ M$^{-1}$cm$^{-1}$nm$^4$ to $1.84 \times 10^{13}$ M$^{-1}$cm$^{-1}$nm$^4$ due to incorporation of nanoclay sheets. Also the presence of nanoclay sheet laponite the intermolecular distance between NK and RhB decreases from 1.4 nm to 0.9 nm. So clay particles play a vital role in concentrating the dyes on their surfaces and thus reducing the intermolecular distance providing a favourable condition for efficient energy transfer. Consequently the energy transfer efficiency increases from 64% to 97% in presence of clay platelets.

It is interesting to mention here that energy transfer has a lot of application in dye lasers. Dye lasers have some limitations as the dye solution used as active medium absorbs energy from the excitation source in a very limited spectral range and so the emission band also has these limitations. If a dye laser has to be used as an ideal source its spectral range needs to be extended. In order to extend the spectral range of operation, mixtures of different dye solutions/dye molecules embedded in solid matrices are being used. The work on energy transfer between different dye molecules in such mixtures in various solvents and solid matrices is, therefore, of great importance. Use of such energy transfer in dye lasers is also helpful in minimizing the photo-quenching effects and thereby, increasing the laser efficiency.

## 8. EFFECT OF NANOCLAY LAPONITE AND IRRADIATION OF MONOCHROMATIC LIGHT ON J-AGGREGATES

J-aggregates have attracted much attention and have been studied intensively by many researchers because of their fascinating optical properties: a red-shifted and narrowed electronic absorption band (J-band), enhanced photoluminescence, and a very small Stokes shift [49]. Nanosized molecular aggregates or J aggregates are interesting phenomena from both the technological and scientific viewpoints, since they are important tools for optical sensitizing, chemical sensing and may be used for information storage [20, 50]. To control the formation of J aggregate is an important target from researcher's point of view.
The effect of the incorporation of clay platelets, laponite, on the J-aggregation of a thiacyanine dye N,N'-dioctadecyl thiacyanine perchlorate (NK) assembled into LB monolayers has been studied [12].

From fig.8 it is found that the absorption spectra of monolayer LB films (without clay) possess prominent band due to H-aggregate at 410 nm, monomer at 433 nm and J-aggregate at 461 nm. In presence of clay platelet laponite the absorption spectra possess only H-aggregate and monomeric band at lower surface pressure, however, only at higher surface pressure J-band appears and dominating over H-aggregate and monomer. The fluorescence spectra of LB monolayer of NK (without clay) possess a broad band at around 450-500 nm regions at lower surface pressure. The LB films deposited at higher surface pressure 30 mN/m the J band is present and it is dominant over the H-aggregate and monomer bands. Thus incorporation of clay platelets makes the Langmuir film more compressible and J-aggregates are formed only in LB films lifted at higher surface pressure of 30 mN/m. J-aggregate of NK molecules are totally absent in the films at lower surface pressure of 10 and 15 mN/m. In one of our recent work we have shown that the J-aggregation of NK molecule which was observed in the LB films lifted at higher surface pressure can be controlled by diluting the NK molecules with OTAB. It was also observed that the J-aggregates of NK decayed to monomer and H-aggregates when the NK-LB film was exposed to a monochromatic light of wavelength 460 nm ($\lambda_{max}$ of J-aggregates) [51].

Fig 9 shows the normalized absorbance, ($A/A_0$) of J-aggregates, monomers and H-aggregates as a function of extent of exposure dose, where A and $A_0$ are the absorbance after and before irradiation, respectively, for the LB films deposited at 30 mN/m surface pressure. It is observed that the normalized absorbance of J-aggregates decreases due to light irradiation and those of monomers and H-aggregates increase. These results are indicative of two possibilities (i) J-aggregates in NK-LB film may decay to monomers and H-aggregates or (ii) NK molecules decay due to light irradiation. In order to clarify these two points, the NK-LB films before and after light irradiation were dissolved in chloroform solution and the absorption spectra of the solutions were measured. It was observed that both the spectra are in good agreement and identical (figure not shown). This confirms that the NK molecules do not decay after light irradiation. Hence, the conclusion is that the J-aggregates decay to monomers and H-aggregates after irradiation of light.

## 9. CONCLUSION AND FUTURE PROSPECTS

Layered inorganics clays are suitable and versatile materials for preparation of nano films. Among them swelling clay minerals or smectites are very unique and they are nanomaterials. They are natural materials, but can also be synthesized in the laboratory. Construction of hybrid clay nano films gives the opportunity to look at clay minerals from a materials scientist's viewpoint. Research should progress along two lines: (1) optimization of the preparation of films towards a particular property (e.g. second harmonic light generation, J-aggregation, FRET etc) or a device (e.g. sensor); (2) fundamental understanding of the film-forming process, of the organization of the molecules, and of the elementary clay mineral platelets and of the stability of the films. Upto now the functionality observed in organo-clay hybrid film are nonlinear optics, two photon absorption, FRET, J-aggregates etc. During the hybrid film formation amphiphilic cation and zwitterionic amphiphilies are being used. As the property of the hybrid films depend on the nature of the amphiphile and their organization. So it would be interesting to try with anionic and neutral amphiphile in order to search new functionality in such films.

There are plenty of opportunities to do original and innovative research. The outcome will certainly be a better knowledge of smectites and their surface properties. This is to the advantage of clay scientists and the clay industry. Possible new industrial applications of clay minerals with high technological value will result from this research.

**ACKNOWLEDGEMENT**

Financial support to carry out this research work through from Department of Atomic Energ, Govt. of India through DAE Young Scientist Research Award (No. 2009/20/37/8/BRNS/3328) has been acknowledged.

**REFERENCES**

[1] Zhou, C.H.; Tong, D.S.; Bao, M.H.; Du, Z.X.; Ge, Z.H.; Li, X.N. Generation and characterization of catalytic nanocomposite materials of highly isolated iron nanoparticles dispersed in clays. *Top. Catal.*, **2006**, *39*, 213-219.
[2] Biswas, S.; Bhattacharjee, D.; Nath, R. K.; Hussain, S. A. Formation of complex Langmuir and Langmuir Blodgett films of water soluble rosebengal. *Journal of Colloid and Interface Science*, **2007**, *311*, 361–367.
[3] Hussain, S. A.; Deb, S.; Biswas, S.; Bhattacharjee, D. Langmuir–Blodgett films of 9-phenyl anthracene molecules incorporated into different matrices. *Spectrochimica Acta Part A*, **2005**, *61*, 2448–2454.
[4] Hussain, S. A.; Islam, Md. N.; Leeman, H.; Bhattacharjee, D.; Aggregation of P-Terphenyl along with PMMA/SA at the Langmuir and Langmuir–Blodgett films. *Surface Review and Letters*, **2008**, *15*, 459-467.
[5] Biswas, S.; Hussain, S. A.; Bhattacharjee, D. Spectroscopic Characterizations of nonamphiphilic 2, 5-bis (5-tert-butyl-benzoxazolyl)-thiophene molecules at the air–water interface and in Langmuir–Blodgett films. *Surface Review and Letters*, **2008**, *15*, 889-896.
[6] Biswas, S.; Hussain, S. A.; Bhattacharjee, D. Orientation of Carbazole molecule in the mixed Langmuir-Blodgett films. *Indian J. Phys.*, **2008**, *82*, 173-177.
[7] Deb, S.; Hussain, S. A.; Biswas, S.; Bhattacharjee, D. Langmuir–Blodgett films of *p*-terphenyl in different matrices: Evidence of dual excimer. *Spectrochimica Acta Part A*, **2007**, *68*, 257–262.
[8] Paul, P. K.; Hussain, S. A.; Bhattacharjee, D. Photophysical characterizations of 2-(4-biphenylyl)-5 phenyl-1,3,4-oxadiazole in restricted geometry. *Journal of Luminescence*, **2008,** *128*, 41–50
[9] Hussain, S. A.; Paul, P. K.; Bhattacharjee, D. Role of various LB parameters on the optical characteristics of mixed Langmuir–Blodgett films. *Journal of Physics and Chemistry of Solids*, **2006**, *67*, 2542–2549.
[10] Hussain, S. A.; Bhattacharjee, D. Langmuir-Blodgett Films and Molecular Electronics. *Modern Physics Letters B*, **2009**, *23*, 3437-3451.
[11] Hussain, S. A.; Chakraborty, S.; Bhattacharjee, D.; Schoonheydt, R. A. Fluorescence Resonance Energy Transfer between organic dyes adsorbed onto nano-clay and Langmuir–Blodgett (LB) films. *Spectrochimica Acta Part A*, **2010**, *75*, 664-670.
[12] Bhattacharjee, D.; Hussain, S. A.; Chakraborty, S.; Schoonheydt, R. A. Effect of nano-clay platelets on the J-aggregation of thiacyanine dye organized in Langmuir-Blodgett films: A spectroscopic investigation. *Spectrochimica Acta Part A*, **2010**, *77*, 232-237.
[13] Hussain, S. A.; Schoonheydt, R. A. Langmuir-Blodgett Monolayers of Cationic Dyes in the Presence and Absence of Clay Mineral Layers: N,N'-Dioctadecyl Thiacyanine, Octadecyl Rhodamine B and Laponite. *Langmuir*, **2010**, *26*, 11870–11877.
[14] Paul, P. K.; Hussain, S. A.; Bhattacharjee, D. Prepartion of ODA-clay hybrid films by Langmuir-Blodgett technique. *Modern Physics Letters B*, **2009**, *23*, 1351-1358.
[15] Hussain, S. A.; Islam, Md. N.; Bhattacharjee, D. Reaction kinetics of organo-clay hybrid films: In-situ IRRAS, FIM and AFM studies. *Journal of Physics and Chemistry of Solids,* **2010**, *71*, 323–328.
[16] Suresh, R.; Borkar, S. N.; Sawant, V. A.; Shende, V. S.; Dimble, S. K. Nanoclay Drug Delivery System. *International Journal of Pharmaceutical Sciences and Nanotechnology,* **2010**, *3*, 901-905.
[17] Garces, J. M.; Moll, D. J. Polymeric Nanocomposites for Automotive Applications. *Adv. Mat.*, **2000**, *12*, 1835-1839.
[18] Batra, M.; Gotam, S.;Dadarwal, P.; Nainwani, R.; Sharma, M. Nano-Clay as Polymer Porosity Reducer: A Review. *Journal of Pharmaceutical Science and Technology*, **2011**, *3*, 709-716.
[19] Carretero, M. I.; Pozo, M. Clay and non-clay minerals in the pharmaceutical and cosmetic industries PartII. Active ingredients. *Applied Clay Science*, **2010**, *47*, 171-181.
[20] Ulman, A. *An Introduction to Ultrathin Organic Films: From Langmuir–Blodgett Films of Self assemblies*, Academic Press, New York, **1991**.
[21] Deb, S.; Biswas, S.; Hussain, S. A.; Bhattacharjee, D. Spectroscopic characterizations of the mixed Langmuir–Blodgett (LB) films of 2, 2'-biquinoline molecules: Evidence of dimer formation. *Chemical Physics Letters*, **2005**, *405*, 323-329.
[22] Hussain, S. A.; Paul, P. K.; Bhattacharjee, D. Role of microenvironment in the mixed Langmuir–Blodgett films. *Journal of Colloid and Interface Science*, **2006**, *299*, 785–790.

[23] Richardsson, T.; Petty, M. C.; Bryce, M. R.; Bloor, D. *Introduction to Molecular Electronics*, Chap. 10 Oxford University Press, New York, **1995**, pp. 221-242.
[24] Hussain, S. A.; Deb, S.; Bhattacharjee, D. Spectroscopic characterizations of non-amphiphilic 2-(4-biphenylyl)-6-phenyl benzoxazole molecules at the air–water interface and in Langmuir–Blodgett films. *Journal of Luminescence*, **2005**, *114*, 197–206.
[25] Islam, Md. N.; Bhattacharjee, D.; Hussain, S. A. Miscibility and Molecular Orientation of Carbazole in Mixed Langmuir and Langmuir-Blodgett Films. *Chin. Phys. Lett.*, **2007**, *24*, 2044-2047.
[26]. Roberts, G. G. *Langmuir Blodgett Films*, Plenum Press, New York, **1990**.
[27]. Petty, M. C. *Langmuir Blodgett Films. An Introduction*, Cambridge University Press, Cambridge, **1996**.
[28] Kotov, N. A.; Meldrum, F. C.; Fendler, J. H.; Tombacz, E.; Dekany, I. Spreading of Clay Organocomplexes on Aqueous Solutions: Construction of Langmuir-Blodgett Clay Organocomplex Films. *Langmuir*, **1994**, *10*, 3797-3804.
[29] Inukai, K.; Hotta, Y.; Taniguchi, M.; Tomura, S.; Yamagishi, A. Formation of a clay monolayer at an air–water interface. *J. Chem. Soc., Chem. Commun.*, **1994**, 959-959.
[30] Umemura, Y.; Yamagishi, A.; Schoonheydt, R.; Persoons, A.; Schryver, F. De. Langmuir−Blodgett Films of a Clay Mineral and Ruthenium(II) Complexes with a Noncentrosymmetric Structure. *J. Am. Chem. Soc.*, **2002**, *124*, 992-997.
[31] Ras, R. H.A.; Németh, J.; Johnston, C. T.; DiMasi, E.; Dékány, I.; Schoonheydt, R. A. Hybrid Langmuir–Blodgett monolayers containing clay minerals: effect of clay concentration and surface charge density on the film formation. *Phys. Chem. Chem. Phys.*, **2004**, *6*, 4174 – 4184.
[32] Umemura, Y.; Onodera, Y.; Yamagishi, A. Layered structure of hybrid films of an alkylammonium cation and a clay mineral as prepared by the Langmuir–Blodgett method. *Thin Solid Films*, **2003**, *426*, 216–220.
[33] Umemura, Y.; Yamagishi, A.; Schoonheydt, R.; Persoons, A.; Schryver, F. De. Formation of hybrid monolayers of alkylammonium cations and a clay mineral at an air-water interface: clay as an inorganic stabilizer for water-soluble amphiphiles. *Thin Solid Films*, **2001**, *388,* 5-8.
[34] Usami, H.; Takagi, K.; Sawaki, Y. Effect of alkylammonium ions on the photochemical behaviors of gamma-stilbazolium aggregates in clay interlayers. *Bull. Chem. Soc. Jpn.*, **1991**, *64*, 3395-3401.
[35] Tapia Esté´vez, M. J.; Lo´pez Arbeloa, F.; Lo´pez Arbeloa, T.; Lo´pez Arbeloa, I.; Schoonheydt, R. A. Spectroscopic study of the adsorption of rhodamine 6G on laponite B for low loading. *Clay Miner.*, **1994**, *29*, 105-113.
[36] Ogawa, M.; Kuroda, K. Photofunctions of Intercalation Compounds. *Chem. Rev.*,**1995**, *95*, 399-438.
[37] Higashi, T.; Miyazaki, S.; Nakamura, S.; Seike, R.; Tani, S.; Hayami, S.; Kawamata, J. Nonlinear optical properties of Langmuir-Blodgett films consisting of metal complexes. *Colloids and Surfaces A: Physicochemical and Engineering Aspects*, **2006**, *284–285*, 161–165.
[38] Kawamata, J.; Seike, R.; Higashi, T.; Inada, Y.; Sasaki, J.; Ogata, Y.; Tani, S.; Yamagishi, A. Clay templating Langmuir-Blodgett films of a non-amphiphilic ruthenium(II) complex. *Colloids and Surfaces A: Physicochemical and Engineering Aspects*, **2006**, *284–285*, 135–139.
[39] Kawamata, J.; Yamaki, H.; Ohshiye, R.; Seike, R.; Tani, S.; Ogata, Y.; Yamagishi, A. Fabrication of hybrid LB films consisting of a smectite clay and a nonamphiphilic chiral ruthenium complex. *Colloids and Surfaces A: Physicochemical and Engineering Aspects*, **2008**, *321*, 65–69.
[40] Ogata, Y.; Kawamata, J.; Chong, C.-H.; Mahikari, M.; Yamagishi, A.; Saito, G. Optical second harmonic generation of zwitterionic molecules aligned on clays. *Molecular Crystals, Molecular Liquids*, **2002**, *376*, 245–250.
[41] Ogata, Y.; Kawamata, J.; Chong, C.-C.; Yamagishi, A.; Saito, G. Strcutural features of a clay film hybridized with a zwitterionic molecule as analyzed by second-harmonic generation behavior. *Clays and Clay Minerals*, **2003a**, *51*, 181–185.
[42] Ogata, Y.; Kawamata, J.; Yamagishi, A.; Chong, C.-H.; Saito, G. A novel film with a noncentrosymmetric molecular alignment of D-p-A zwitterionic molecules fabricated at an air-water interface. *Synthetic Metals*, **2003b**, *133–134*, 671–672.
[43] Kawamata, J.; Suzuki, Y.; Tenma, Y. Fabrication of clay mineral-dye composites as nonlinear optical materials. *Philosophical Magazine*, **2010**, *90*, 2519–1527.
[44] Suzuki, Y.; Hizakawa, S.; Sakamoto, Y.; Kawamata, J.; Kamada, K.; Ohta, K. Hybrid films consisting of a clay and a diacetylenic two-photon absorption dye. *Clays and Clay Minerals*, **2008**, *56*, 487–493.
[45] Kamada, K.; Tamamura, Y.; Ueno, K.; Ohta, K.; Misawa. Enhanced two-photon absorption of chromophores confined in two-dimensional nanospace. *Journal of Physical Chemistry C*, **2007**, *111*, 11193–11198.
[46] Forster, T. H. *Naturforsh*, **1949**, *4a*, 321–327.
[47] Forster, T. H. *Diss. Faraday Soc.*, **1959**, *27*, 7–71.
[48] Brouard, D.; Viger, M. L.; Guillermo Bracamonte, A.; Boudreau, D. Label-Free Biosensing Based on Multilayer Fluorescent Nanocomposites and a Cationic Polymeric Transducer. *ACS Nano*, **2011**, *5*, 1888–1896.

[49] Kobayashi, T. (Ed.), *J-Aggregates*, World Scientific: Singapore, **1996**.
[50] Khun, H.; Mobius, D.; Bucher, H. *Techniques of Chemistry, in: A. Weissberger*; B.W. Rossiter, Eds.; New York, **1973**; vol. *1*, Part IIIB.
[51] Hussain, S. A.; Dey, D.; Chakraborty, S.; Bhattacharjee, D. J-aggregates of thiacyanine dye organized in LB films: effect of irradiation of light. *Journal of Luminescence*, **2011**, *131*, 1655-1660.

Figure Caption:

Fig 1. Basic 2:1 clay minerals

Fig. 2. (a) The general chemical structure of a LB compatible molecule with a carboxylic acid head and an arbitrary tail (b) A typical LB film deposition instrument installed in our laboratory.

Fig. 3. Schematic of the Langmuir–Blodgett (LB) film deposition process. (a) The amphiphile is dissolved in an volatile organic solvent and subsequently spread at the air–water interface. The solvent evaporates and a monolayer of the amphiphile remains at the air–water interface (b). The monolayer at the air–water interface can be further manipulated by means of a movable barrier allowing control of the surface pressure and molecular packing in the monolayer (c). The Langmuir monolayer can be transferred onto solid substrate to form mono- or multilayer LB films. (d) Transfer of monolayer on to a hydrophilic substrate occurred during an up-stroke (e) and via a down-stroke on to a hydrophobic surface. By repeating number of up-stroke and down-stroke desired number of layered LB films may be obtained. LB monolayer onto substrate are also shown.

Fig. 4. Schematic of surface pressure - area per molecule ($\pi$-A) isotherm showing different phases of monolayer at air-water interface [10].

Fig. 5 Schematic diagram showing organo-clay hybrid film formation using LB technique (a) Spreading of amphiphilic cation onto the LB trough filled up with clay dispersion. (b) Barrier is compressed to form compact Langmuir monolayer after waiting 30 mins. To complete the adsorption process. (c) Floating film is being transferred onto solid substrate during upstroke.

Fig. 6. AFM images of some organo-clay hybrid films using different types of clay

Fig. 7. Fluorescence spectra of pure NK, RhB and their mixture (50:50 volume ratio) in presence and absence of nano clay sheet laponite [11].

Fig. 8. (a) Absorption and fluorescence spectra of monolayer NK LB films without clay lifted at different surface pressures viz. 10, 15 and 30 mN/m. (b) Absorption and fluorescence spectra of monolayer NK LB films with clay lifted at different surface pressures viz. 10, 15 and 30 mN/m [12].

Fig. 9. Normalized absorbance ($A/A_0$) of J-aggregates, monomers and H-aggregates as a function of exposure extent for a NK-LB monolayer film deposited at 30 mN/m surface pressure [51].

Figures:

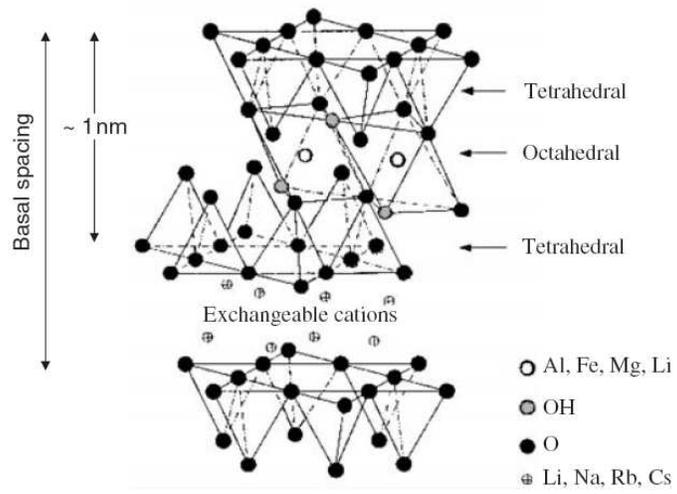

Fig.1

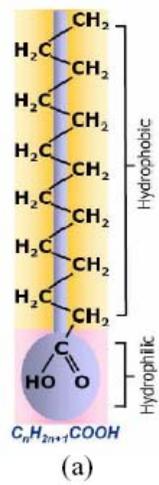
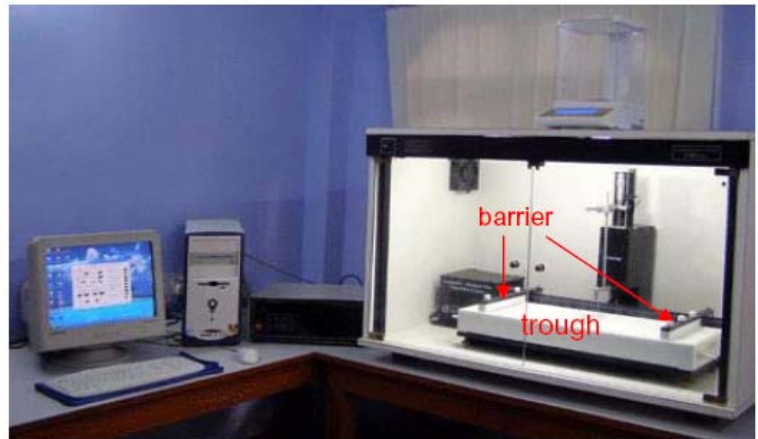

Fig.2

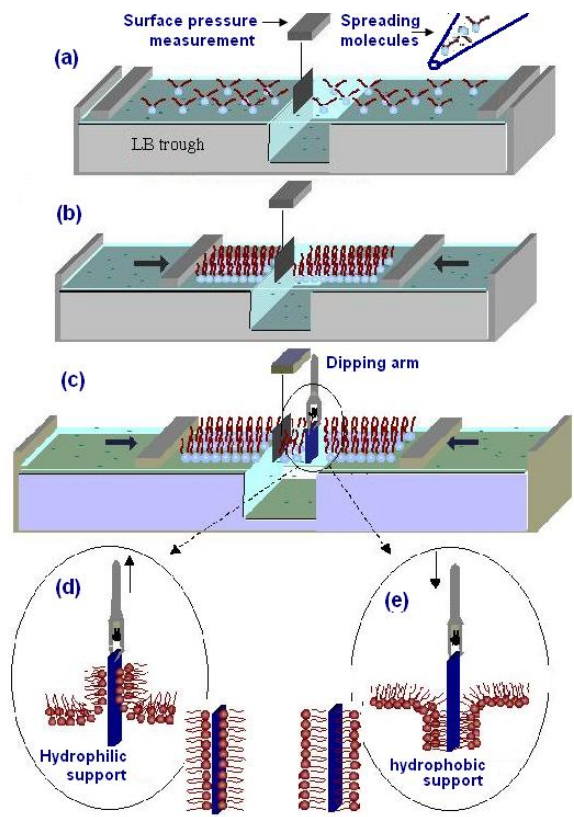

Fig. 3

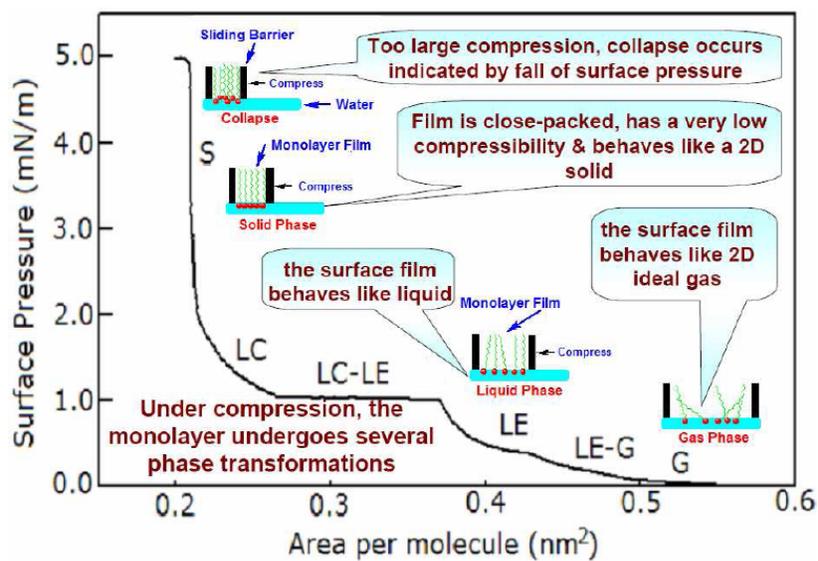

Fig.4

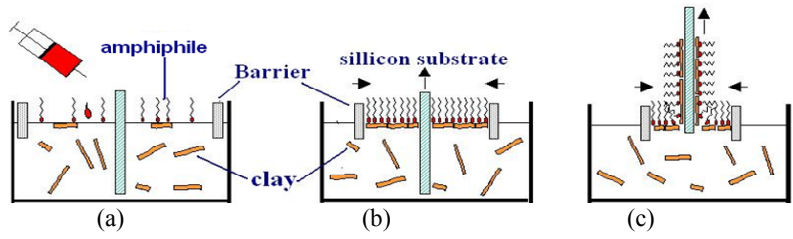

Fig.5

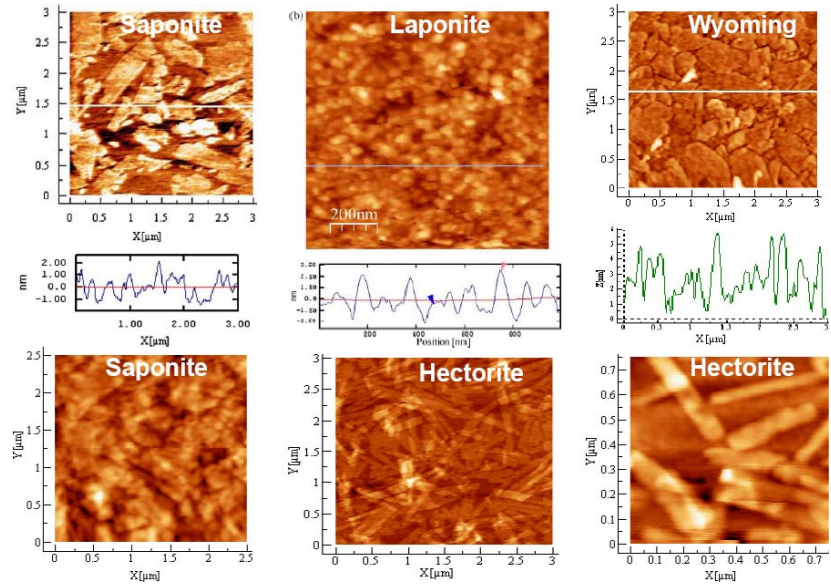

Fig.6

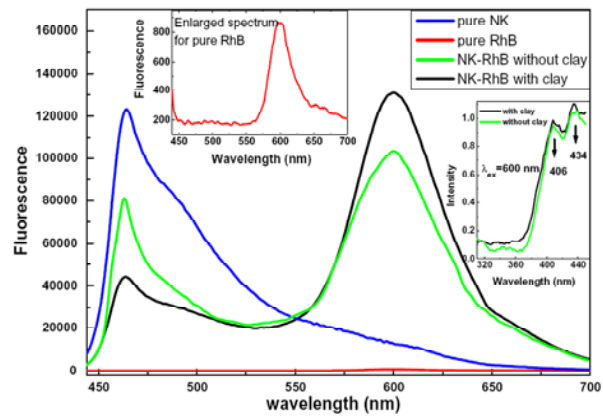

Fig.7

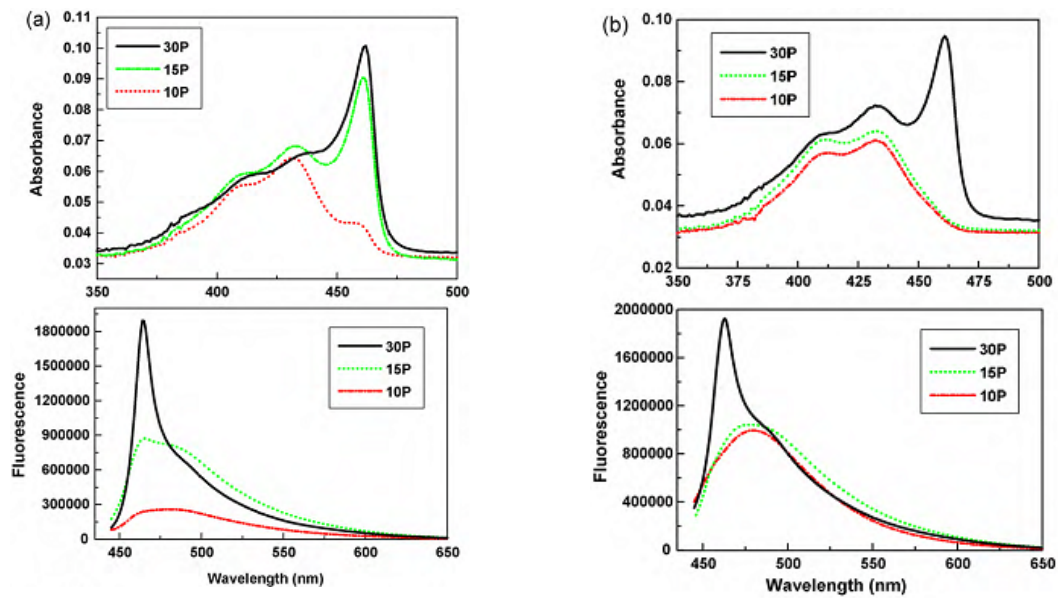

Fig. 8

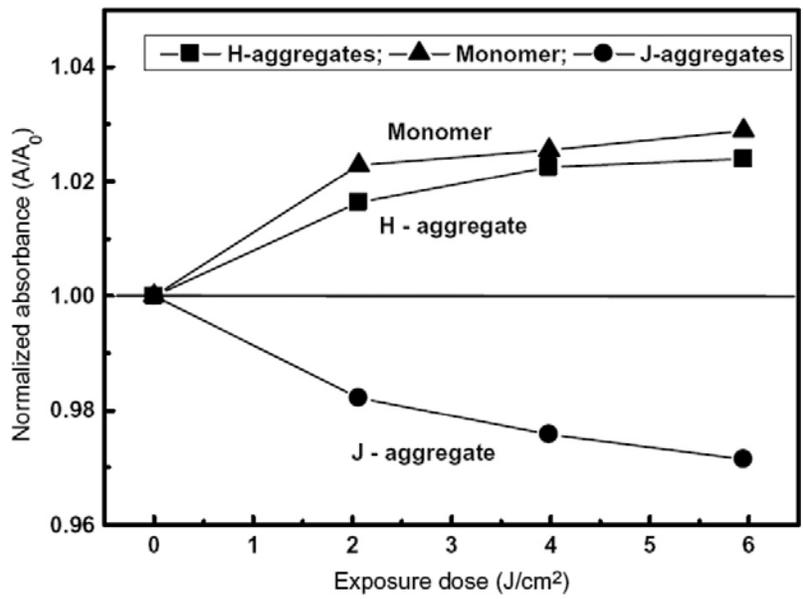

Fig.9

16